\documentstyle[12pt]{article}
\begin{document}

\title{{\Large {\bf An Ising Spin-$S$ Model on Generalized Recursive
Lattice }}}
\author{N.S. Ananikian\thanks{e-mail: ananik@jerewan1.yerphi.am}\\
{\normalsize Yerevan Physics Institute, Alikhanian Br.2, 375036 Yerevan, Armenia}\\
N.Sh. Izmailian \thanks{e-mail: izmailan@phys.sinica.edu.tw} \thanks{Permanent address: Yerevan Physics Institute, Alikhanian Br.2, 375036 Yerevan, Armenia}\\
{\normalsize Institute of Physics, Academia Sinica, Taipei, Taiwan 
11529, Republic of China}\\
K.A. Oganessyan \thanks{e-mail: kogan@hermes.lnf.infn.it} \thanks{Permanent 
address: Yerevan Physics Institute, Alikhanian Br.2, 375036 Yerevan, 
Armenia}\\
{\normalsize LNF-INFN, C.P. 13, Enrico Fermi 40, Frascati, Italy}} 
\maketitle

\begin{abstract}
The Ising spin - $S$ model on recursive $p$ - polygonal 
structures in the external magnetic field is considered and the general 
form of the free energy and magnetization for arbitrary spin is derived. 
The exact relation between the free energies on infinite entire 
tree and on its infinite "interior" is obtained.   
\end{abstract}

\newpage

\begin{center}
1. INTRODUCTION
\end{center}

Many theoretical problems of statistical mechanics, solid state physics,  
gauge models, polymers, etc., can be solved exactly on recursive 
structures, like the Bethe \cite{A1,A2,A3,A4} and the Husimi \cite{A5,A6,A7} 
lattices. Moreover, it has been argued recently \cite{A8} that, in some case, 
the Bethe lattice calculations are more reliable than mean-field 
calculations. In connection with this, the knowledge of the exact form of the 
free energy and magnetization is very actual for locating phase transitions  
and for analytical investigations of physical phenomenon. The main problem is 
how to calculate the free energy on the infinite "interior" of the infinite 
entire lattice. Namely, how get rid of the surface (with nonzero density of 
sites) contributions which can allow to strange results and rather different 
from those on regular lattices \cite{A9}. In our previous  papers the 
relation between the free energies on Bethe and Cayley trees for spin-1 Ising 
model \cite{A3} and on Husimi and corresponded to it Cayley trees for 
multisite Ising model \cite{A7} are obtained. In that approach the free 
energy derived from the recursion relations and then the differentiation of 
the free energy functionals over the external magnetic field allowed to the 
exact relations. Recently, a method of computing the bulk free energy for any 
model on infinite "interior" tree is presented \cite{A8}.      

In this paper we have generalized the approach of the surface independent 
free energy calculation \cite{A3,A7} for an Ising spin - $S$ model in an 
external magnetic field on generalized recursive lattice consisted of $p$ - 
polygons ($p$ is the number of edges of the polygon). 

\begin{center}
2. THE MODEL
\end{center}

The recursive tree consisted of $p$ - polygons is characterized by $p$, the 
number of edges (the number of sites) of the polygon ($p=2$ - usual Bethe 
lattice, $p=3$ - Husimi lattice) and by $q$, the number of p-polygons which 
go out from each site (Fig.1). 

These infinite dimensional recursive lattices cannot be embedded in
any lattice belonging to an Euclidean space of finite dimension. However, 
they can be embedded in two-dimensional space of constant negative
curvature (the hyperbolic or Lobachevsky plan {\bf H2}) and study their
metric properties \cite{A10}.

Let us define the Ising spin-$S$ model in an external  magnetic field 
by Hamiltonian 
\begin{equation}
\label{R1}
-{\beta}H=\sum_{p-poligon}{H_1({\sigma}_{i}^{(1)},{\sigma}_{i}^{(2)},
\cdots {\sigma}_{i}^{(p)})}+\sum_i{H_2({\sigma}_i)},
\end{equation}
where ${\sigma}_i$ take values $(-S,-S+1,...,S-1,S)$, ${\sigma}_{i}^{(k)}$ -
spin around polygon, the first sum goes over all $p$ - polygons and the
second over all sites.. $H_1({\sigma}_i^{(1)},{\sigma}_i^{(2)},...,{\sigma}_i^{(p)})$ - includes all possible nearest - neighbor pair interactions, as well 
multisite interactions between spins belongs the same polygon and $H_2({\sigma}_i)=\sum_{\mu =1}^{2S}h_{\mu}{\sigma}_i^{\mu}$ - includes all possible single 
- ion interactions.

The partition function will have the form 
\begin{equation}
\label{R2}
Z=\sum_{\{ \sigma \}} 
\exp{\left\{ \sum_{p-poligon}{H_1({\sigma}_{i}^{(1)},{\sigma}_{i}^{(2)}, \cdots
{\sigma}_{i}^{(p)})}+\sum_i{H_2({\sigma}_i)} \right \}},
\end{equation}

The advantage of the recursive tree is that for models formulated on it 
exact recursion relations can be derived. When this tree is cut apart at the
central site 0, it separates into $q$ identical branches, each of
which contains $(q-1)(p-1)$ branches. Then the partition function can 
be written as follows:
\begin{equation}
\label{R3}
Z=\sum_{\sigma_0}\exp{\left\{H_2({\sigma}_0)\right\}}
g_N^q({\sigma}_0),
\end{equation}
where ${\sigma}_0$ is a spin in central site, $N$ is the number of 
generations, and $g_N({\sigma}_0)$ is the partition function of a 
branch. Each branch, in turn, can be cut along any site of the 
1th-generation which is nearest to the central site. The expression for 
$g_N({\sigma}_0)$ can therefore be written in the form 
\begin{equation}
\label{R4}
g_N({\sigma}_0)=\sum_{\{\sigma_1^{(j)}\}}\exp{\left 
\{H_1({\sigma}_{0}, {\sigma}_{1}^{(1)},\cdots {\sigma}_{1}^{(p-1)})+
\sum_{i=1}^{p-1} H_2({\sigma}_{1}^{(i)}) \right\}}
g_{N-1}^{q-1}({\sigma}_1^{(1)}) \cdots
g_{N-1}^{q-1}({\sigma}_1^{(p-1)}),
\end{equation}
where ${\sigma_1^{(j)}}$ are the spins of first polygon except the central 
site. Consequently, we will have $2S+1$ recursion relations for the 
$g_N({\sigma_0})$,  where ${\sigma_0}$ take values $(-S, -S+1,....,S-1, S)$.
After dividing each of the recursion relations on a recursion relation for
$g_N(S)$, we will have $2S$ recursion relations for $x_N({\sigma_0})$ 
\begin{equation}
\label{R5}
x_N(\sigma_0)=\frac{\sum_{\{\sigma_1^{(j)}\}}\exp\left
\{H_1(\sigma_{0},\sigma_1^{(1)},\cdots\sigma_1^{(p-1)})+
\sum_{i=1}^{p-1} H_2(\sigma_1^{(i)})\right\}
x_{N-1}^{q-1}(\sigma_1^{(1)})\cdots
x_{N-1}^{q-1}(\sigma_1^{(p-1)})}{\sum_{\{\sigma_1^{(j)}\}}
\exp\left\{H_1(S,\sigma_1^{(1)},\cdots
\sigma_1^{(p-1)})+\sum_{i=1}^{p-1}H_2(\sigma_1^{(i)})\right\}
x_{N-1}^{q-1}(\sigma_1^{(1)})\cdots
x_{N-1}^{q-1}(\sigma_1^{(p-1)})},
\end{equation}
where 
\begin{equation}
\label{R6}
x_N(\sigma_0)=g_N(\sigma_0)/g_N(S),
\end{equation}
and an equation for $g_N(S)$
\begin{equation}
\label{R7}
g_N(S)=g_{N-1}^{(q-1)(p-1)}(S)\Psi_{N-1},
\end{equation}
where 
\begin{equation}
\label{R8}
\Psi_{N-1}=\sum_{\{\sigma_1^{(j)}\}} \exp\left \{H_1(S,
{\sigma}_{1}^{(1)},\cdots {\sigma}_{1}^{(p-1)})+
\sum_{i=1}^{p-1} H_2({\sigma}_{1}^{(i)}) \right \}
x_{N-1}^{q-1}({\sigma}_1^{(1)}) \cdots
x_{N-1}^{q-1}({\sigma}_1^{(p-1)}).
\end{equation}
Since the right-hand side of Eq. (\ref{R5}) is bounded in $x_N$, it follows 
that $x_N$ is finite for $N \to \infty$.

Through this $x_N(\sigma)$ one can express the density 
$m_{\mu}=\langle {\sigma}_0^{\mu}\rangle$ of central site (the symbol 
$\langle ... \rangle$ denotes the thermal average), where $\mu$ takes values 
from $1$ to $2S$:
\begin{equation}
\label{R9}
m_{\mu}= \langle {\sigma }_0^{\mu} \rangle =\frac{\sum_{\sigma_0}{\sigma }_0^{\mu}\exp{\left\{H_2({\sigma}_0)\right\}
x_N^q({\sigma}_0)}}
{\sum_{\sigma_0}\exp{\left\{H_2({\sigma}_0)\right\}
x_N^q({\sigma}_0)}}
\end{equation}
and other thermodynamic parameters. So we can say that the $x_N(\sigma)$ in 
the thermodynamic limit $(N \to \infty)$ determine the states of the system.

Using the Eq. (\ref{R6}), the free energy on the Cayley tree can be 
written as follows 
\begin{equation}
\label{R10}
-{\beta}F_N^{Cayley}=q\ln{g_N(S)}+\ln(\Phi_N),
\end{equation} 
where 
\begin{equation}
\label{R11}
\Phi_N=\sum_{\sigma_0}\exp
H_2({\sigma}_0)x_N^q({\sigma}_0).
\end{equation}
By substituting in the right-hand side of Eq. (\ref{R10}) the expression of 
$g_N(S)$ given by 
Eq. (\ref{R7}), we obtain 
the recursion relation for the free energy on the generalized Cayley tree
\begin{equation}
\label{R12}
-{\beta}F_N^{Cayley}= -(q-1)(p-1){\beta}F_{N-1}^{Cayley} + q\ln(\Psi_{N-1}) -
(q-1)(p-1)\ln(\Phi_{N-1}) + \ln(\Phi_N) 
\end{equation}
By repeating this recursion procedure $n$ times one can obtain
\begin{equation}
\label{R13}
-\beta F_N^{Cayley}=-[(q-1)(p-1)]^n{\beta}F_{N-n}^{Cayley}-\beta F_{nN},
\end{equation}
where
\begin{equation}
\label{R14}
-\beta F_{nN}= q\sum_{k=1}^n[(q-1)(p-1)]^{k-1}\ln(\Psi_{N-k}) - 
[(q-1)(p-1)]^n\ln(\Phi_{N-n}) + \ln(\Phi_N).
\end{equation}

Therefore we obtain the general form of the free energy for Ising 
spin-$S$ model on generalized Cayley tree   
\begin{equation}
\label{R15}
-{\beta}F_N^{Cayley}= q\sum_{k=0}^{N-1}[(q-1)(p-1)]^{N-k-1}\ln(\Psi_k) + 
\ln(\Phi_N) 
\end{equation}
where $\Psi_k$ and $\Phi_N$ are given by Eqs. (\ref{R8}) and (\ref{R11}). 

As mentioned in the Introduction the above obtained free energy  
gives rise to unusual behavior due to surface sites (in particular, it 
depends on surface conditions) \cite{A9}. To overcome this problem we here 
consider only local properties of sites deep within the graph (i.e. 
infinitely far from the boundary in the limit $N \to \infty$). Indeed,  
we will consider only contribution to $Z_N$ from the Bethe lattice. 

Let us consider the case, when the series of solution 
of recursion relations given by Eq. (\ref{R5}) 
converge to a stable point at $N \to \infty$. In this case 
$\lim{x_{N-n}{({\sigma})}}=x({\sigma})$ for all finite $n$ and the 
recursion equations become 
\begin{equation}
\label{R16}
x(\sigma_0) = \frac{\sum_{\{\sigma_1^{(j)}\}}\exp\left
\{H_1(\sigma_{0},\sigma_1^{(1)},\cdots\sigma_1^{(p-1)})+
\sum_{i=1}^{p-1} H_2(\sigma_1^{(i)})\right\}
x^{q-1}(\sigma_1^{(1)})\cdots
x^{q-1}(\sigma_1^{(p-1)})}{\sum_{\{\sigma_1^{(j)}\}}
\exp\left\{H_1(S,\sigma_1^{(1)},\cdots
\sigma_1^{(p-1)})+\sum_{i=1}^{p-1}H_2(\sigma_1^{(i)})\right\}
x^{q-1}(\sigma_1^{(1)})\cdots
x^{q-1}(\sigma_1^{(p-1)})}. 
\end{equation}

Then the  Eq. (\ref{R14})  will take the form 
$$
-\beta F_n = \lim_{N \to \infty}{(-\beta F_{nN})}
$$
\begin{equation}
\label{R17}
= q\frac{(q-1)^n(p-1)^n-1}{(q-1)(p-1)-1}\ln(\Psi) -
[(q-1)^n(p-1)^n-1]\ln(\Phi),
\end{equation}
where $\Psi=\lim_{N \to \infty}{{\Psi}_{N-k}}$ and 
$\Phi=\lim_{N \to \infty}{{\Phi}_{N-k}}$
for all finite $k$ and 
\begin{equation}
\label{R18}
\Psi=\sum_{\{\sigma_1^{(j)}\}} \exp\left \{H_1(S,
{\sigma}_{1}^{(1)},\cdots {\sigma}_{1}^{(p-1)})+
\sum_{i=1}^{p-1} H_2({\sigma}_{1}^{(i)}) \right \}
x^{q-1}({\sigma}_1^{(1)}) \cdots
x^{q-1}({\sigma}_1^{(p-1)}),
\end{equation}
\begin{equation}
\label{R19}
\Phi=\sum_{\sigma_0}\exp
H_2({\sigma}_0)x^q({\sigma}_0).
\end{equation}

Let us now prove that Eq. (\ref{R17}) gives the free energy of the spin - 
$S$ Ising model on the generalized Bethe lattice consist of $n$ generations.  

First, let calculate the total number of the sites of the generalized Bethe 
lattice consist of $n$ generations. By definition, the Bethe lattice is the 
union of equivalent sites and therefore it is easy to get the relation 
between the numbers of bonds and sites. As from each site go out $q$ bonds 
and each bond contains two sites then $N_{bond} = {q \over 2} N_{site}$. On 
the other hand, the number of bonds of the Cayley tree or the Bethe lattice 
consist of $n$ generations is 
$$
N_{bond} = q+q(q-1)+q(q-1)^2+\cdots+q(q-1)^{n-1} = q\frac{(q-1)^n-1}{q-2}. 
$$
So, the total number of sites of the Bethe lattice consist of $n$ generations 
is 
$$
N_{site} = {2 \over q} N_{bond} = 2 \frac{(q-1)^n-1}{q-2}.
$$         
Above we calculate the the number of sites in the case when $p=2$. It is easy 
to generalize this result for arbitrary $p$. Note, only, that in case when 
$p>2$ from each site go out $2q$ bonds and the relation between the numbers 
of bonds and sites will looks like 
$$
N_{bond} = {2q \over 2} N_{site} = q N_{site},
$$ 
and the number of bonds for arbitrary $p$ will have the following form  
$$
N_{bond} = qp+qp(q-1)(p-1)+\cdots+qp(q-1)^{n-1}(p-1)^{n-1}= 
         qp \frac{(q-1)^n(p-1)^n-1}{(q-1)(p-1)-1}. 
$$
Thus, the total number of sites of the generalized Bethe lattice   
\footnote
{One can reach to the same result in a different way as well. 
Indeed, one of the ways to fulfill the equivalent of the sites of the 
Bethe lattice is to put a periodical boundary condition on Cayley tree. Then 
the number of sites of the last shell of Cayley tree will be equal to 
$q(q-1)^{n-1}(p-1)^{n-1}$, whereas the number of sites of the corresponding 
Bethe lattice is $(q-1)^{n-1}(p-1)^{n-1}$. One can see that the number of 
sites on the boundary shall of the generalized Bethe lattice is $q$ times 
less then the corresponding number on the generalized Cayley lattice. 
Consequently
$$
N_{site}^{Bethe} = 1+q(p-1)+q(q-1)(p-1)^2+\cdots+q(q-1)^{n-2}(p-1)^{n-1}
$$
$$
+ (q-1)^{n-1}(p-1)^n = p\frac{(q-1)^n(p-1)^n-1}{(q-1)(p-1)-1}. 
$$
} 
consist of $n$ generations is  
\begin{equation}
\label{R20}
N_{site}^{Bethe} = p\frac{(q-1)^n(p-1)^n-1}{(q-1)(p-1)-1}.
\end{equation}

Now we can write the explicit expression for the free energy per site 
for the spin - $S$ Ising model on the generalized Bethe lattice 
\begin{equation}
\label{R21}
-\beta f_{Bethe} = -\frac{\beta F_n}{N_{site}^{Bethe}} = \frac{q}{p}\ln(\Psi) 
- \frac{(q-1)(p-1)-1}{p}\ln(\Phi),
\end{equation}
where $\Psi$ and $\Phi$ are given by Eqs. (\ref{R18}) and  (\ref{R19}) 
respectively. 

We wish to mention here that in our earlier paper \cite{A3}, 
we had noted that this functional gives the exact form of the free energy for 
spin-1/2 Ising model on the Bethe lattice (p=2) \cite{A11}. Then we checked 
the 
correctness of it for a spin-1 Ising model on the Bethe lattice (p=2) by 
differentiation over the external magnetic field. Recently, the 
correctness of that functional is confirmed for multisite 
antiferomagnetic spin-1/2 Ising model on the Husimi tree (p=3) as well 
\cite{A7}. 

Below an effort is made to prove that this statement is correct for 
arbitrary spin - S Ising model on the generalized Bethe lattice.

With this aim in view, let us differentiate this free energy functional with 
respect to the external field $h_{\mu}$. Without 
loss of generality, we will consider, for simplicity, the case $p=2$. 
For arbitrary $p$, the proof will be accomplished in the same way. 

After a little algebra, we obtained 
\begin{eqnarray}
\label{R22}
-\frac{\delta}{\delta h_{\mu}}(\beta f_{Bethe}) &=& \frac{\sum_{\sigma_0}
{\sigma }_0^{\mu}\exp{\left\{H_2({\sigma}_0)\right\}x^q({\sigma}_0)}}
{\sum_{\sigma_0}\exp{\left\{H_2({\sigma}_0)\right\}x^q({\sigma}_0)}}
\nonumber \\
&+&\frac{q}{2}
\frac{\sum_{{\sigma}_0}\exp{\left\{H_2(\sigma_0)\right\}}x^{q-1}({\sigma}_0)
{\bf T}(\sigma_0)}{\sum_{\sigma,{\sigma}_0}\exp{\left\{H_1(S,\sigma)+ H_2(\sigma)+H_2({\sigma}_0)\right\}}x^{q-1}(\sigma)x^{q-1}({\sigma}_0)}.
\end{eqnarray} 
Here 
 
$$
{\bf T}(\sigma_0) = \sum_{\sigma}\exp{\left\{H_2(\sigma)\right\}}
x^{q-1}(\sigma)
$$
\begin{eqnarray}
\label{R23}                                                 
\times \left\{[{\sigma}^k+(q-1)\frac{x'(\sigma)}{x(\sigma)}]
[x({\sigma}_0)\exp{\left\{H_1(S,\sigma)\right\}}-\exp{\left\{H_1(\sigma_0,\sigma)\right\}}]+x'(\sigma_0)
\exp{\left\{H_1(S,\sigma)\right\}}\right\}, 
\end{eqnarray}
where 
$$
x'(\sigma) \equiv \frac{\delta}{\delta h_{\mu}}x(\sigma).
$$
It is easy to check that
\begin{equation}
\label{R24}
{\bf T}({\sigma}_0)=\frac{\delta}{\delta h_{\mu}}
\left\{
\sum_{\sigma}\exp{\left\{H_2(\sigma)\right\}}x^{q-1}(\sigma)
\left [
x({\sigma}_0)\exp\left\{H_1(S,\sigma)\right\} - 
\exp\left\{H_1(\sigma_0,\sigma)\right\}\right ]
\right\}.
\end{equation}
After substituting in the right-hand side of Eq. (\ref{R24}) the expression 
of $x(\sigma_0)$ 
(Eq. (\ref{R16})), which in the case $p = 2$ is
\begin{equation}
\label{R25}
x(\sigma_0)=\frac{\sum_{\sigma}\exp\left
\{H_1(\sigma_0,\sigma)+H_2(\sigma)\right\}
x^{q-1}(\sigma)}{\sum_{\sigma}
\exp\left\{H_1(S,\sigma)+H_2(\sigma)\right\}
x^{q-1}(\sigma)},
\end{equation}
we will obtain
$$
{\bf T}({\sigma}_0)=0
$$
and consequently
\begin{equation}
\label{R26}
-\frac{\delta}{\delta h_{\mu}}(\beta f_{Bethe}) = m_{\mu} = \frac{\sum_{\sigma_0}
{\sigma }_0^{\mu}\exp{\left\{H_2({\sigma}_0)\right\}x^q({\sigma}_0)}}
{\sum_{\sigma_0}\exp{\left\{H_2({\sigma}_0)\right\}x^q({\sigma}_0)}}. 
\end{equation}
 
Thus, the expression given by Eq. (\ref{R21}) is the exact free energy 
functional per site for an Ising spin-$S$ model on the generalized Bethe 
lattice. Consequently, the equation (Eq. (\ref{R13})) \footnote{More recently, 
an elegant geometrical interpretation of this relation is presented in Ref. 
\cite{A8}. The key idea of that method is the construction of a union trees 
with coordination number $q$ from initial tree with the same coordination 
number.} is the exact relation between the free energies on generalized 
Cayley tree and generalized Bethe lattice for a spin - $S$ Ising model. It is 
obvious that this relation is model independent and depends only on the 
structure of recursive tree. 
   
\begin{center}
3. CONCLUSION
\end{center}

In this paper we have considered an Ising spin - $S$ model on recursive $p$ - 
polygonal structures in the external magnetic field and derive the general 
form of the free energy and magnetization for arbitrary spin. We have 
obtained  the exact relation between the free energies on infinite entire 
tree and on its infinite "interior" and show that it is model independent and 
depends only on structure of recursive tree. The advantage of approach 
presented in this paper is that it gives possibility to obtain the exact 
expression for the surface independent free energy for width class of spin 
systems on generalized recursive lattices. This approach should be applicable 
for gauge models on generalized multi-plaquette recursive structures as 
well..

\begin{center}
4. Acknowledgment
\end{center}
The authors are grateful to R. Flume, and S. Ruffo for useful 
discussions. This work was partly supported by the Grant-211-5291 YPI of 
the German Bundesministerium fur Forshung and Technologie and by 
INTAS-96-960.

\newpage

\begin{center}
5. Figure Caption
\end{center}

Fig.1 The recursive tree with $p=4$ and $q=2$. The numbers $0, 1, 2$ and $3$ 
denote the central site, the sites of 1th, 2nd and 3rd generations, 
respectively.  

\end{document}